\documentclass[12pt]{iopart}
\usepackage{iopams}

\input epsf.sty

\newlength{\TZ}
\newcommand{\BEQ}{\begin{equation}}     
\newcommand{\BEA}{\begin{eqnarray}}
\newcommand{\EEQ}{\end{equation}}       
\newcommand{\EEA}{\end{eqnarray}}
\newcommand{\eps}{\varepsilon}          
\newcommand{\D}{{\rm d}}                
\newcommand{\demi}{\frac{1}{2}}         
\newcommand{\ket}[1]{\left|#1\right\rangle}  

\renewcommand{\vec}[1]{\boldsymbol{#1}} 

\catcode`\@=11
\def\numberbysection{\@addtoreset{equation}{section}
        \def\theequation{\thesection.\arabic{equation}}}

\begin{document}

\title[Ageing: bosonic contact and pair-contact processes]{Ageing without 
detailed balance in the bosonic contact and pair-contact processes: exact
results}

\author{Florian Baumann$^{1,2}$, Malte Henkel$^2$, Michel Pleimling$^1$ and
Jean Richert$^3$}
\address{$^1$Institut f\"ur Theoretische Physik I, 
Universit\"at Erlangen-N\"urnberg, \\
Staudtstra{\ss}e 7B3, D -- 91058 Erlangen, Germany}
\address{$^2$Laboratoire de Physique des 
Mat\'eriaux,\footnote{Laboratoire associ\'e au CNRS UMR 7556} 
Universit\'e Henri Poincar\'e Nancy I, \\ 
B.P. 239, F -- 54506 Vand{\oe}uvre l\`es Nancy Cedex, France}
\address{$^3$Laboratoire de Physique Th\'eorique,\footnote{Laboratoire 
associ\'e au CNRS UMR 7085} Universit\'e Louis Pasteur Strasbourg, \\
3, rue de l'universit\'e, F -- 67084 Strasbourg Cedex, France} 

\begin{abstract}
Ageing in systems without detailed balance is studied in the exactly 
solvable bosonic contact process and the critical bosonic pair-contact
process. The two-time correlation function and the two-time
response function are explicitly found. In the ageing regime, the dynamical
scaling of these is analyzed and exact results for the ageing exponents and
the scaling functions are derived. For the critical bosonic pair-contact
process the autocorrelation and autoresponse exponents agree but the ageing 
exponents $a$ and $b$ are shown to be distinct. 
\end{abstract}

\submitto{\JPA}
\maketitle

\setcounter{footnote}{0}

\section{Introduction}

Systems brought rapidly out of an initial state into a region in parameter
space which is characterized by several competing stationary states may undergo
ageing behaviour. Dynamical scaling and universality was first noticed in
the mechanical properties of several glass-forming systems rapidly quenched
into their glassy phase \cite{Stru78} and has since been found and studied
intensively in a large variety of systems relaxing towards an equilibrium 
state, see \cite{Bray94,Cate00,Cugl02,Godr02,Cris03,Henk04,Cala04} for recent 
reviews. In what follows we shall restrict to systems without any macroscopic
conservation laws and to systems without frustrations, conditions which are
paradigmatically met in simple ferromagnets with dynamics which satisfy
detailed balance. 

Convenient tools for the study of ageing behaviour in such
systems \cite{Cugl94b}
are the two-time autocorrelation and autoresponse functions, 
which in the {\it ageing regime} $t,s\gg 1$ and $t-s\gg 1$ are expected 
to show the scaling behaviour
\BEA
C(t,s) &=& \langle \phi(t) \phi(s) \rangle ~~ \sim s^{-b} f_C(t/s) 
\label{gl:C} \\
R(t,s) &=& \left.\frac{\delta\langle\phi(t)\rangle}{\delta h(s)}\right|_{h=0}
\sim s^{-1-a} f_{R}(t/s) 
\label{gl:R}
\EEA
Here $\phi(t)$ is the order parameter and $h(s)$ is the conjugate field, $t$
is called the observation time and $s$ the waiting time. 
For large arguments $y\to \infty$, one generically expects
\BEQ
\label{gl:lambda}
f_C(y) \sim y^{-\lambda_C/z} \;\; , \;\;
f_R(y) \sim y^{-\lambda_R/z}
\EEQ
where $\lambda_C$ and $\lambda_R$, respectively, are known as autocorrelation
\cite{Fish88,Huse89} and autoresponse exponents \cite{Pico02}. While the 
ageing exponents $a$ and $b$ can be expressed in terms of the dynamical 
exponent $z$ and equilibrium exponents, the exponents $\lambda_{C,R}$ are 
independent of these but related to the so-called initial
slip exponents \cite{Jans92}. For critical quenches
\BEQ \label{gl:ab}
a= b = \frac{2\beta}{\nu z} \;\; , \;\; \mbox{\rm at $T=T_{\rm c}$}
\EEQ
(where $\beta$ and $\nu$ are the usual equilibrium critical exponents) 
is a consequence of the fluctuation-dissipation theorem and of time-translation
invariance in the scale-invariant equilibrium steady-state. 

Turning to the scaling functions, it has been suggested \cite{Henk01,Henk02} 
that their form could be determined from the requirement of covariance under
so-called {\em local} scale-transformations which are constructed from the
requirement of including the special conformal transformations 
$t\mapsto (\alpha t+\beta)/(\gamma t+\delta)$ with $\alpha\delta-\beta\gamma=1$
in time. For the response function this leads to \cite{Henk01,Henk02,Henk05b}
\BEQ \label{gl:Rf}
R(t,s) = s^{-1-a} f_R(t/s) \;\; , \;\; f_R(y) = f_0\, y^{1+a'-\lambda_R/z}
(y-1)^{-1-a'}
\EEQ
where $a'$ is a new independent exponent and $f_0$ is a normalization
constant. This form (or equivalently an integrated response) describes very
well all available numerical data for ferromagnetic systems quenched to
$T<T_{\rm c}$ \cite{Henk01,Henk02,Henk05a} and was shown to be 
exact in several exactly solvable models, see \cite{Godr02} and refs. therein. In 
these systems, $z=2$ \cite{Bray94}. On the other hand, for quenches to 
$T=T_{\rm c}$, one has in general $z\ne 2$, but the agreement of (\ref{gl:Rf})
with numerical data is still almost perfect
\cite{Henk01,Abri04b,Plei04}. However, a second-order 
$\eps$-expansion calculation in the O($n$)-symmetric $\phi^4$-field theory
produced a small but systematic correction with respect to 
(\ref{gl:Rf}) \cite{Cala04} and
there is support of this finding from a numerical study of $R(t,s)$ in
momentum space \cite{Plei04b}. 
While in most systems studied so far $a=a'$ holds true, 
several models with $a\ne a'$ are also known \cite{Henk05b}. 

The approach of local scale-invariance uses the dynamical symmetry of 
deterministic equations, whereas the influence of noise is essential in the
understanding of non-equilibrium dynamics. However, considering a description
of the ageing system in terms of a Langevin equation, it can be shown that 
{\em if} the deterministic (i.e. noiseless) part satisfies local 
scale-invariance, then (i) the response function $R(t,s)$ is noise-independent
and (ii) the autocorrelator $C(t,s)$ can be reduced to noise-less three- and 
four-point response functions \cite{Pico04}. A further extension of local
scale-transformation to include the diffusion constant as a new dynamical 
variable then permits to determine $C(t,s)$ and the result was found to be
in good agreement with simulational data in the $2D$ Ising model quenched to
$T<T_{\rm c}$ \cite{Henk04b}.

An important ingredient in the ageing studies discussed so far is the
assumption of detailed balance for the dynamics. This begs the question what
might happen if that condition is relaxed. Indeed, numerical studies of 
the contact process, the simplest system of that kind, gave the following
results \cite{Enss04,Rama04}:
\begin{enumerate}
\item Dynamical scaling and ageing only occur {\em at} the critical point. 
This is expected since both inside the active and the inactive phases 
there merely is a single stable stationary state. 
\item At criticality, the scaling forms eqs.~(\ref{gl:C},\ref{gl:R}) hold true
for the (connected) autocorrelator and the response function, but with the
scaling relation
\BEQ \label{abverschieden}
a+1 = b = \frac{2\beta}{\nu_{\perp} z},
\EEQ
in contrast to eq.~(\ref{gl:ab}), with $\beta$ and $\nu_{\perp}$ being now 
standard steady-state exponents. 
\end{enumerate}
In order to get a better understanding of these results, it would be helpful
to study the ageing behaviour in exactly solvable but non-trivial models 
without detailed balance. Remarkably, it has been realized by Houchmandzadeh
\cite{Houc02} and by Paessens and Sch\"utz \cite{Paes04a} that the bosonic
versions of the contact process and of the critical 
pair-contact process (where an arbitrary number
of particles are allowed on each lattice site) are exactly solvable, at 
least to the extent that the dynamical scaling behaviour of
equal-time correlators can be
analysed exactly \cite{Houc02,Paes04a,Paes04b}. Here we extend
their work by means of an exact calculation of the two-time correlation and 
response functions for the bosonic contact process and the critical bosonic 
pair-contact process. In section~2, we define the models and write down the
closed systems of equations of motion for the correlation and response 
functions. We also recall the existing results on the single-time correlators
\cite{Houc02,Paes04a}. In section~3, we discuss the bosonic contact process
and in section~4, we describe our results for the critical pair-contact 
process. In section~5 we give the results for the two-time response functions. 
As we shall see, the critical bosonic pair-contact process provides 
a further example of a model where $a$ and $b$ are different. 
In section~6 we conclude. A detailed discussion of local scale-invariance in 
these models will be presented in a sequel paper.

\section{The models}

Consider the following stochastic process: on an infinite $d$-dimensional
hypercubic lattice particles move diffusively with rate $D$ in each spatial
direction. Each site may contain an arbitrary non-negative number of particles. 
Furthermore, on any given site the following reactions for the particles
$A$ are allowed
\BEA
m A \longrightarrow (m+k)A \;\; ; \;\; \mbox{\rm with rate $\mu$} 
\nonumber \\
p A \longrightarrow (p-\ell )A \;\; ; \;\; \mbox{\rm with rate $\lambda$}
\label{gl:rates}
\EEA
It is to be understood that on a given site, out of any set of $m$
particles $k$ additional particles are created with rate $\mu$ and $\ell$ 
particles are destroyed out of any set of $p\geq \ell$ particles with rate
$\lambda$. Diffusion applies on single particles. We shall be concerned with
two special cases:
\begin{enumerate}
\item the {\em bosonic contact process}, where $p=m=1$, hence $\ell=1$. 
The value of $k$ is unimportant and will be fixed to $k=1$ as well. 
\item the {\em bosonic pair-contact process}, where $p=m=2$. 
\end{enumerate}

While the bosonic contact process arose from a study on the origin of 
clustering in biology \cite{Houc02}, the bosonic pair-contact process as
defined here \cite{Paes04a} is an offshoot of a continuing debate
about the critical behaviour of the diffusive pair-contact process (PCPD), see
\cite{Henk04c} for a recent review. Initially, this model was 
introduced \cite{Howa97} in an attempt to understand the meaning
of `imaginary' versus `real' noise but the associated field theory turned 
out to be unrenormalizable \cite{Howa97,Jans04}. 
A lattice version (with the `fermionic' constraint of not more than one
particle per site) of the model contains the reactions $2A\to\emptyset$ and
$2A\to 3A$ together with single-particle diffusion 
$A\emptyset\leftrightarrow\emptyset A$ and was first studied numerically 
in \cite{Carl01}. An intense debate on the universality class of this 
model followed, see \cite{Henk04c}, and several 
mutually exclusive conclusions on the critical behaviour continue to be drawn, 
see \cite{Jans04,Bark03,Kock03,Park04,Szol04,Hinr05} for recent work.
The bosonic pair-contact process has a dynamic exponent $z=2$ \cite{Paes04a} 
and is hence distinct from the PCPD where $z<2$. 
Its study will not so much shed light on any open question
concerning the PCPD but it should rather be viewed as a non-trivial 
example of an exactly solvable non-equilibrium many-body system to be studied 
in its own right. 

The master equation is written in a quantum hamiltonian
formulation as $\partial_t \ket{P(t)} = - H \ket{P(t)}$ 
\cite{Doi76,Schu00} where $\ket{P(t)}$ is the time-dependent state vector 
and the hamiltonian $H$ can be
expressed in terms of annihilation and creation operators
$a(\vec{x})$ and $a^\dag(\vec{x})$. We define also the particle
number operator as $n(\vec{x}) = a^\dag(\vec{x})
a(\vec{x})$. Then the Hamiltonian of the model (\ref{gl:rates})
reads \cite{Paes04a}
\BEA
\label{gl:hamiltonian}
\hspace{-1truecm} H &\hspace{-0.5truecm}=& - D \sum_{r=1}^{d} \sum_{\vec{x}} 
\left[ a(\vec{x})a^{\dag}(\vec{x}+\vec{e}_r) + 
a^{\dag}(\vec{x})a(\vec{x}+\vec{e}_r) - 2n(\vec{x}) \right] 
\nonumber \\
& & - \lambda \sum_{\vec{x}} \left[ 
\left(a^{\dag}(\vec{x})\right)^{p-\ell} \left( a(\vec{x})\right)^p 
- \prod_{i=1}^{p} \left( n(\vec{x})-i+1\right) \right]
\label{eq:qH} \\
& & - \mu \sum_{\vec{x}} \left[ 
\left(a^{\dag}(\vec{x})\right)^{m+k} \left( a(\vec{x})\right)^m 
- \prod_{i=1}^{m} \left( n(\vec{x})-i+1\right) \right] 
- \sum_{\vec{x}} h(\vec{x},t) a^{\dag}(\vec{x})~~~~
\nonumber 
\EEA
where $\vec{e}_r$ is the $r^{\rm th}$ unit vector. For later use in the 
calculation of response functions we have also added an external field 
which describes the spontaneous creation of a single particle 
$\emptyset\to A$ with a site-dependent rate $h=h(\vec{x},t)$ on
the site $\vec{x}$. 

\begin{figure}[htb] 
  \vspace{0.5cm}
  \centerline{\epsfxsize=5.0in\epsfclipon\epsfbox
  {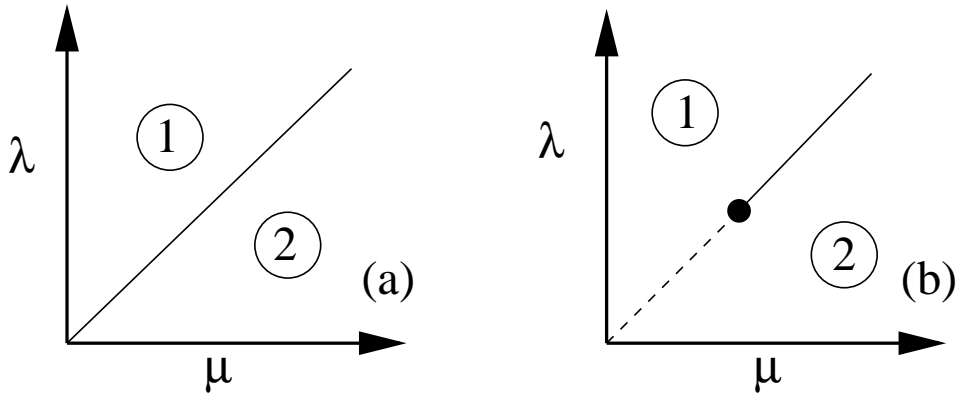}
  }
  \caption[Phasendiagramm]{Schematic phase-diagrams for $D\ne 0$ of (a) 
  the bosonic contact process and the bosonic pair-contact process in $d\leq 2$ 
  dimensions and (b)
  the bosonic pair-contact process in $d>2$ dimensions. The active region 1, 
  where $\lim_{t \rightarrow \infty} \rho(\vec{x},t)= 0$, is separated
  by the critical line eq.~(\ref{gl:Krit}) from the absorbing region 2, 
  where $\rho(\vec{x},t)\to\infty$ as $t\to\infty$. On the
  critical line $\rho(t):=\int\!\D\vec{x}\, \rho(\vec{x},t)$ remains constant. 
  By varying $\alpha$ one moves along the critical line.
  Along the critical line, one may have clustering (full lines in (a) and (b)), 
  but in the bosonic pair-contact process with $d>2$ the steady-state may also 
  be homogeneous (broken line in (b)). These two regimes are separated by a
  multicritical point.}
  \label{fig:Abb0}
\end{figure}

Single-time observables $g(\vec{x},t)$ can be obtained from
the time-independent quantities $g(\vec{x})$ by switching to the
Heisenberg picture. They satisfy the usual Heisenberg 
equation of motion, from which the differential equations
for the desired quantities can be obtained. The space-time-dependent
particle-density $\rho(\vec{x},t) := \langle
a^{\dag}(\vec{x},t) a(\vec{x},t) \rangle = \langle
a(\vec{x},t) \rangle $ satisfies
\BEQ
\frac{\partial}{\partial t} \left\langle a(\vec{x},t)\right\rangle 
= D \Delta_{\vec{x}} \left\langle a(\vec{x},t)\right\rangle 
-\lambda\ell \left\langle a(\vec{x},t)^p\right\rangle
+\mu k \left\langle a(\vec{x},t)^m\right\rangle + h(\vec{x},t)
\EEQ
where we have used the short-hand
\BEQ
\Delta_{\vec{x}} f(\vec{x}) := \sum_{r=1}^{d}  
\left( f(\vec{x}-\vec{e}_r,t) + f(\vec{x}+\vec{e}_r,t)
-2f(\vec{x},t) \right)
\EEQ
and similar equations hold for the equal-time two-point correlation functions,
see \cite{Paes04a}. It turns out that for the bosonic contact process
$p=m=1$ these equations close for arbitrary values of the rates. On the
other hand, for the bosonic pair-contact process where $p=m=2$ a closed
system of equations is only found along the critical line given by \cite{Paes04a}
\BEQ \label{gl:Krit}
\ell \lambda = \mu k.
\EEQ
This line separates an active phase with a formally infinite particle-density
in the steady-state from an absorbing phase where 
the steady-state particle-density vanishes, see figure~\ref{fig:Abb0} 
for the schematic phase-diagrams. In what follows,
the essential control parameter is
\BEQ \label{gl:def_controlparameter}
\alpha := {\mu k (k+\ell)}/(2D)
\EEQ
The physical nature of this transition becomes apparent when 
equal-time correlations are studied \cite{Houc02,Paes04a} and can be formulated
in terms of a {\sl clustering transition}. By clustering we mean that particles
accumulate on very few lattice sites while the other ones remain empty. Now,
for the bosonic contact process, the behaviour along the critical line is
independent of $\alpha$. If $d\leq 2$, there is always clustering, while
there is no clustering for $d>2$. On the other hand, in the bosonic 
pair-contact process, there is on the critical line a multicritical 
point at $\alpha=\alpha_C$, with 
\BEQ \label{gl:def_alphaC}
\alpha_C =\alpha_C(d) = \frac{1}{2A_1} \;\; , \;\;
A_1 := \int_0^{\infty} \!\D u\, \left( e^{-4u} I_0(4u)\right)^d
\EEQ
and where $I_0(u)$ is a modified Bessel function \cite{Abra65}, such 
that clustering occurs for $\alpha>\alpha_C$ only and with a more or less
homogeneous state for $\alpha\leq \alpha_C$. Specific values are
$\alpha_C(3) \approx 3.99$ and $\alpha_C(4) \approx 6.45$ and
$\lim_{d\,\searrow\, 2}\alpha_C(d)=0$. 
We are interested in studying the
impact of this clustering transition on the two-time correlations and linear
responses. 

In order to obtain the equations of motion of the two-time correlator, the 
time-ordering of the operators $a(\vec{x},t)$ must be taken in account. {}From 
the Hamiltonian eq.~(\ref{eq:qH}) without an external field $h$,
we get the following equations of 
motion for the two-time correlator, after rescaling the 
times $t\mapsto t/(2D)$, $s\mapsto s/(2D)$, and for $t>s$,
\cite{Glauber} (for a detailed computation, see \cite{Baum05a})
\newpage \typeout{ *** hier ist ein Seitenvorschub ! *** }
\BEA
\label{gl:eqn_of_motion} 
& & ~~~~ \frac{\partial}{\partial t} \left\langle a(\vec{x},t) a(\vec{y},s)\right\rangle 
\\
&=& \frac{1}{2} \Delta_{\vec{x}} \left\langle a(\vec{x},t) 
a(\vec{y},s)\right\rangle
-\frac{\lambda\ell}{2D} \left\langle a(\vec{x},t)^p a(\vec{y},s)\right\rangle  
+\frac{\mu k}{2D} \left\langle a(\vec{x},t)^m a(\vec{y},s)\right\rangle
\nonumber 
\EEA
which we are going to study in the next sections. 

\section{The bosonic contact process} 

For the bosonic contact process, we have $p=m=1$, hence also
$\ell=k=1$. We first consider the critical case $\lambda\ell=\mu k$. We 
shall assume throughout that spatial translation-invariance holds and use 
the notation
\BEQ
F(\vec{r};t,s) := \left\langle a(\vec{x},t)
a(\vec{x}+\vec{r},s)\right\rangle. 
\EEQ
Then $F$ satisfies a diffusion equation which is solved in a standard way by
Fourier transforms. It is easy to see that the solution of the 
equations of motion (\ref{gl:eqn_of_motion}) involves the single-time
correlator $F(\vec{r},t):=F(\vec{r};t,t)$ which satisfies
the equation of motion, after the usual rescaling $t\mapsto
t/(2D)$ \cite[eq.~(10)]{Paes04a},
\BEQ
\frac{\partial}{\partial t}F(\vec{r},t) = \Delta_{\vec{r}} F(\vec{r},t)
+ \alpha \rho_0 \delta_{\vec{r},\vec{0}}
\EEQ
and the parameter $\alpha$ was defined in (\ref{gl:def_controlparameter}). 
As initial conditions, we shall use throughout the Poisson distribution 
$F(\vec{r},0)= \rho_0^2$. Hence one arrives at the 
following expression of our main quantity of interest, 
the connected correlator\footnote{In 
\cite{Enss04,Rama04} this same quantity was denoted by $\Gamma(t,s)$ which
we avoid here in order not to create confusion with the incomplete 
gamma function \cite{Abra65}.} 
\BEQ
\label{gl:gdef}
G(\vec{r};t,s) := F(\vec{r};t,s)- \rho_0^2 =  
\alpha\rho_0 \int_0^{s} \!\D\tau\: b\left(\vec{r},\frac{1}{2}(t+s)-\tau\right)
\EEQ
where ($I_r(t)$ being a modified Bessel function) 
\BEQ \label{gl:bessel}
b(\vec{r},t) = e^{-2dt} I_{r_1}(2t) \ldots I_{r_d}(2t).
\EEQ

We evaluate this expression in two cases 
\begin{itemize}
  \item \underline{$\vec {r} = 0$, $t$ and $s$ in the ageing regime: } \\
    In this case both $s$ and $t-s$ are large, so
    that we can use the asymptotic behaviour 
    $I_0(t)\simeq (2\pi t)^{-1/2} e^t$ for $t$ large \cite{Grad80} 
    for the expression $b(\vec{0},\frac{1}{2}(t+s) - \tau)$ under the integral 
    in (\ref{gl:gdef}). 
    We have to distinguish the cases $d > 2$ , $d = 2$ and
    $d< 2$. For $d > 2$ we obtain 
    \BEA
    \label{gl:resultm1ar}
    G(\vec{0},t,s) &\simeq& \frac{\alpha \rho_0}{(4
    \pi)^{\frac{d}{2}}} \int_0^s \!\D\tau \left(\frac{1}{2}(t+s) -
    \tau\right)^{-\frac{d}{2}} \nonumber \\ &=& \frac{\alpha \rho_0}{(4
    \pi)^{\frac{d}{2}} (\frac{d}{2}-1)} \left(
    \left(\frac{t-s}{2}\right)^{-\frac{d}{2}+1} -
    \left(\frac{t+s}{2}\right)^{-\frac{d}{2}+1} \right).
    \EEA
    By analogy with eqs.~(\ref{gl:C})
    and (\ref{gl:lambda}), we expect the scaling behaviour
    $G(t,s) := G(\vec{0};t,s)=s^{-b} f_G(t/s)$. We read off the value 
    $b = \frac{d}{2}-1$ and the scaling function
    \BEQ
    \label{gl:scalingfunction}
    f_G(y) = \frac{\alpha \rho_0}{2
    (2\pi)^\frac{d}{2}(\frac{d}{2}-1)}\left(
    (y-1)^{-\frac{d}{2}+1} - (y+1)^{-\frac{d}{2}+1} \right).
    \EEQ
    {}From the expected asymptotics $f_G(y)\sim y^{-\lambda_G/z}$ for $y\gg 1$,
    we obtain
    \BEQ
    \lambda_G = d
    \EEQ
    as can be seen from the asymptotic development of
    (\ref{gl:scalingfunction}) and where we anticipated that the dynamical
    exponent $z=2$, see also \cite{Houc02} and below.\\
    
    For $d = 2$ the integral in (\ref{gl:resultm1ar})  gives a
    different result. We find
    \BEQ \label{gl:Bos2d}
    G(t,s) = f_G(t/s) \;\; , \;\; 
    f_G(y) = \frac{\alpha \rho_0}{2
    (2\pi)^\frac{d}{2}}\ln \left( \frac{y+1}{y-1} \right)
    \EEQ
    and we have the exponents $b=0$ and $\lambda_G=2$. The logarithmic 
    divergence of the single-time correlator \cite{Houc02} reflects itself
    here in the logarithmic form of the scaling function. \\
     
    Finally, for $d < 2$ the same computation as for $d > 2$ goes through. 
    Now, the exponent $b=\frac{d}{2}-1$ is {\em negative} which means that
    the two-time autocorrelator diverges, in agreement with the
    earlier results for the equal-time correlators in $1D$ \cite{Houc02}. 
  \item \underline{$\vec{r}$-dependence for $s,t-s \gg 1$} \\
    We use the asymptotic expression, valid for $u\gg 1$ and $\vec{r}^2/u$  
    fixed 
    \BEQ
    \label{gl:asymptotic_beh2}
    e^{-d z}I_{r_1}(u)\cdot \ldots \cdot I_{r_d}(u)\simeq \frac{1}{(2 \pi
    u)^{\frac{d}{2}}} \exp \left(- \frac{\vec{r}^2}{2 u}
    \right)
    \EEQ
    which yields for arbitrary dimension $d$, when introduced into
    (\ref{gl:gdef}),
    \BEA
    \label{gl:resultm1_r}
    \hspace{-2.0truecm}G(\vec{r};t,s) &\simeq& 
    \frac{\alpha \rho_0}{(4\pi)^{\frac{d}{2}}} 
    \left(\frac{\vec{r}^2}{4}\right)^{-(\frac{d}{2}-1)} 
    \left[ \Gamma \left(\frac{d}{2}-1, \frac{1}{2}\frac{\vec{r}^2}{t+s}\right)
    -\Gamma \left(\frac{d}{2}-1, \frac{1}{2}\frac{\vec{r}^2}{t-s}\right)\right]
    \EEA
    The incomplete Gamma
    function $\Gamma(\kappa,x)$ is defined by \cite{Abra65}
    \BEQ
    \Gamma(\kappa,x) := \int_x^\infty \!\D t\: e^{-t}
    t^{\kappa-1}
    \EEQ
    and has the following asymptotic behaviour for large or small
    arguments
    \BEQ
    \label{gl:asymptotic_gamma}
    \Gamma(\kappa,x) \stackrel{|x| \gg 1}{\approx}
    x^{\kappa-1} e^{-x} ,\qquad \Gamma(\kappa,x)
    \stackrel{0<x\ll1}{\approx} \Gamma(\kappa) -
    \frac{x^\kappa}{\kappa}.
    \EEQ
    In the limit where both $s$ and $t-s$ become large, we
    recover eq.~(\ref{gl:resultm1ar}) as it should be. Furthermore, we
    explicitly see that the dynamical exponent $z=2$. 
\end{itemize}

\begin{figure}[h]
  \centerline{\epsfxsize=5.0in\epsfclipon\epsfbox
  {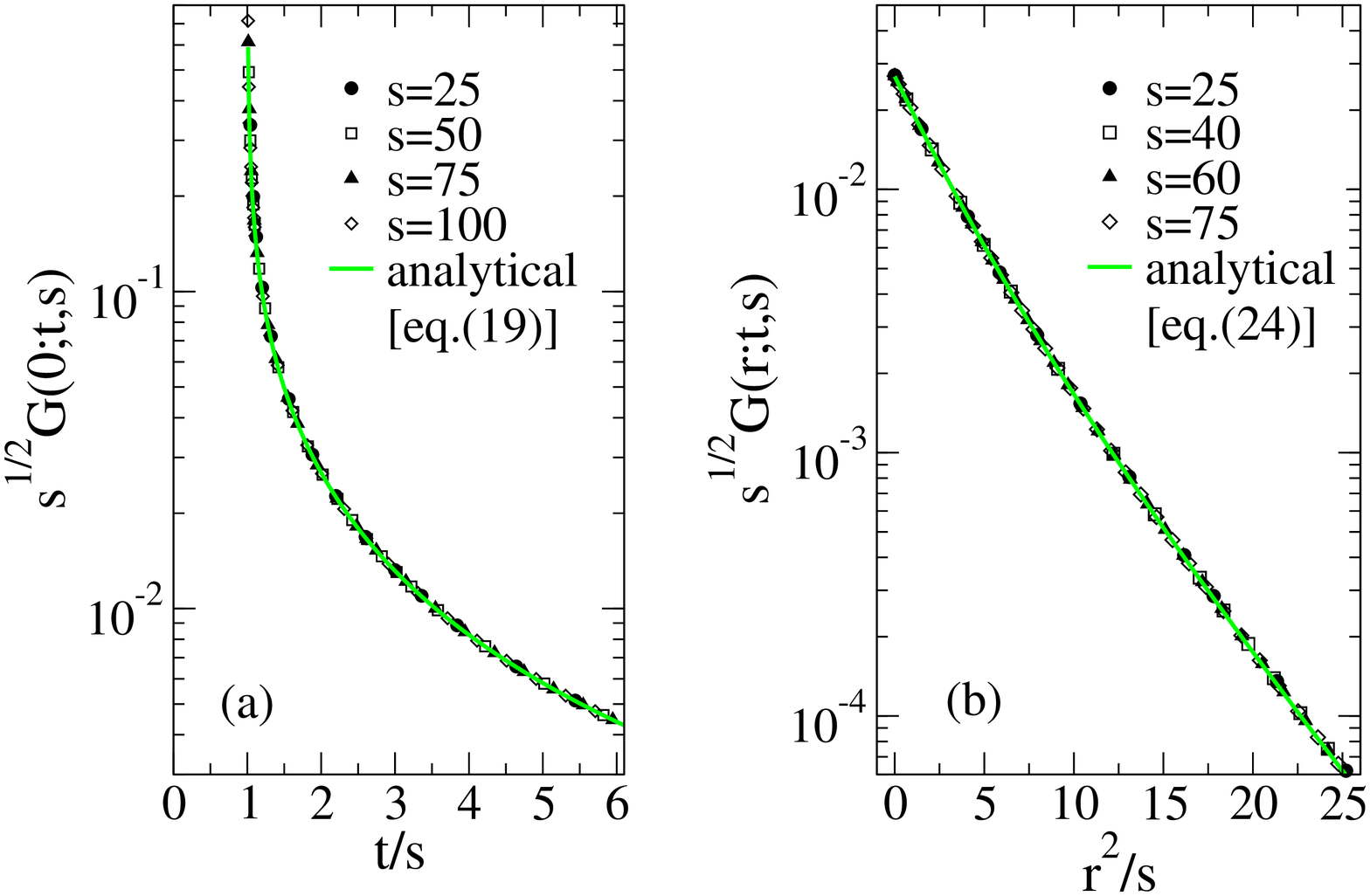}
  }\caption{Scaling plots of (a) the autocorrelation
  function $G(0;t,s)$ and (b) the space-dependent
  correlation function $G(\vec{r};t,s)$ for the critical bosonic
  contact process in three dimensions
  with $\alpha \rho_0 = 1$. In (b), the
  value of $y = {t}/{s}=2$ was used.} 
  \label{fig:p1m1}
\end{figure} 

For illustration, we have also evaluated the integral 
(\ref{gl:gdef}) numerically (with $\alpha \rho_0 = 1$).  
In Figure~\ref{fig:p1m1}a we compare the numerical
results, for several values of $s$ in three dimensions, with the
analytical result eq.~(\ref{gl:resultm1ar}). We see that already for
quite small values of $s$ one has a nice data collapse which confirms the
expected scaling behaviour. Furthermore, the agreement with the 
analytically calculated scaling function is perfect. 
In figure~\ref{fig:p1m1}b we display the dependence on  $\vec{r}$, 
evaluated along the line $\vec{r} = (r,0,\ldots,0)$. 
Again, the expected scaling
behaviour is also confirmed and the curves agree with the
analytical expression eq.~(\ref{gl:resultm1_r}).

In the non-critical case, we have for the density, after
rescaling $t \rightarrow {t}/(2 D)$
\BEQ
\label{gl:equation_of_motion_delta}
\frac{\partial}{\partial t} \rho(\vec{x},t) = \frac{1}{2}
\Delta_{\vec{x}} \rho(\vec{x},t) + \frac{1}{2} \eta
\rho(\vec{x},t) \qquad \mbox{with} \qquad  \eta := \frac{\mu k -
\lambda \ell}{D}
\EEQ
This is easily solved and yields 
\BEQ
\label{gl:density}
\rho(\vec{x},t) = \rho_0 e^{\frac{1}{2} \eta t}
\EEQ
if we choose again a homogenous initial distribution with mean density 
$\rho_0$. Depending on whether particle creation or
annihilation dominates the density increases or decreases
exponentially. Next, for the single-time correlator $F(\vec{r},t)$ 
we use \cite[eqs. (7,8)]{Paes04a} which can be written after rescaling 
as (recall $\ell=1$)
\BEQ
\frac{\partial}{\partial t} F(\vec{r},t) = \Delta
F(\vec{r},t) + \eta F(\vec{r},t) + \alpha
\delta_{\vec{r},0}\, \rho(t) 
\EEQ
which is easily solved by introducing the particle-density (\ref{gl:density})
and performing a Fourier transform. This in turn allows to solve the
equation of motion (\ref{gl:eqn_of_motion}) for the two-time correlator 
and we find 
\BEQ
F(\vec{r};t,s) = \rho_0^2 e^{\frac{1}{2}\eta (t+s)} +
\alpha \rho_0 e^{\frac{1}{2} \eta (t+s)} \int_0^s \!\D\tau\:
e^{-\frac{1}{2} \eta \tau} b\left(\vec{r};\frac{1}{2}(t+s) -
\tau \right).
\EEQ
We consider the case $\vec{r} = 0$ and $t$ and $s$ in the
ageing regime. As before, we use 
the asymptotic expression for $b(\vec{0},\frac{1}{2}(t+s) - \tau)$ and find
for the connected autocorrelator 
\BEA
\label{gl:calc_delta} 
\hspace{-1.5truecm}G(\vec{0};t,s) &\hspace{-0.1truecm}~:=& 
\hspace{-0.0truecm}F(\vec{0};t,s) - \rho_0^2
e^{\frac{1}{2}\eta (t+s)} \nonumber \\ 
&\hspace{-0.9truecm}=& \hspace{-0.4truecm}\frac{\alpha \rho_0
e^{\frac{1}{4} \eta (t+s)}}{(4 \pi)^{{d}/{2}}}
\left[\Gamma \left(-\frac{d}{2}+1,-\frac{\eta}{4} \, (t+s)\,
\right) -
\Gamma \left(-\frac{d}{2}+1,-\frac{\eta}{4} \, (t-s)
\right)\right].
\EEA
Using the asymptotic behaviour eq.~(\ref{gl:asymptotic_gamma}) 
for the Gamma-function for large arguments we obtain
\BEQ
\hspace{-1.6truecm} G(\vec{0};t,s) =  
- \frac{2\alpha \rho_0}{(2 \pi)^{{d}/{2}} \, \eta}
\left[ {(t-s)}^{-d/2} {\exp\left(\frac{\eta}{2} t\right)}
- {(t+s)}^{-d/2} \exp\left(\frac{\eta}{2}(t+s)\right)
\right]. 
\EEQ
If $\eta$ is positive, then particle-creation outweighs
particle-annihilation. The second term dominates and leads to an
exponential divergence. On the other hand, if $\eta$ is negative, 
the first term involving $e^{\eta t/2}$ is the dominant one. 
At first sight, these results appear curious, since the leading exponential
behaviour merely depends on $t+s$ and $t$, respectively, and not on $t-s$, 
as might have been anticipated. 

A similar result had already been found in the inactive phase of the 
ordinary contact process \cite{Enss04,Rama04} and we can understand the
present result along similar lines. Consider the limits 
where $|\eta|\to\infty$, such that diffusion plays virtually no role in
comparison with the creation or annihilation processes. Then merely the
creation and annihilation processes on a single site need to be considered. 
Correlators are given in terms of conditional probabilities and we now
consider the two cases $\eta>0$ and $\eta<0$. 
First, for $\eta<0$, annihilation dominates and at late times there
are only few particles left in the system. This is the same situation as in
the inactive phase of the ordinary contact process. Then $G(\vec{0};t,s)$ 
can only be non-vanishing if at time $s$ a particle was present and it
should only depend on $t$. On the other hand, for $\eta>0$, the 
particle-density diverges exponentially and the number of possible reactions
is conditioned by the density at time $s$, proportional to
$e^{\eta s/2}$,
hence the dependence on $t+s$. Finally, the power-law prefactors relate to the
diffusion between different sites. 

\section{The bosonic critical pair-contact process}
For the bosonic pair-contact process, we have $p=m=2$.
The system (\ref{gl:eqn_of_motion}) of differential equations closes only 
for the critical case, i.e. for $\lambda\ell=\mu k$, and we shall
restrict to this situation throughout. At criticality, the values of
$\ell$ and $k$ do not influence the scaling behaviour. 
It was shown in \cite{Paes04a} that in dimensions $d>2$ there is a 
phase transition along the critical line and we must therefore distinguish 
three cases, according to whether the reduced control parameter 
\BEQ
\alpha' := \frac{\alpha-\alpha_C}{\alpha_C}.
\EEQ
is negative, zero, or positive and where $\alpha$ was defined in 
(\ref{gl:def_controlparameter}) and $\alpha_C$ in (\ref{gl:def_alphaC}). 
For $d\leq 2$ one is always in the situation $\alpha'>0$. 
We recall the known results for the single-time autocorrelator $F(\vec{0},t)$
which for large times behaves as \cite{Paes04a}
\begin{itemize}
  \item \underline{$\alpha < \alpha_C:$}  
    \BEQ
    \label{gl:asymptotic1}
    F(\vec{0},t) \stackrel{t
    \rightarrow \infty}{\approx} - \frac{\rho_0^2}{\alpha'}.
    \EEQ
  \item \underline{$\alpha = \alpha_C:$}  
    \renewcommand{\arraystretch}{1.6}
    \BEQ
    \label{gl:asymptotic3}
    F(\vec{0},t) \stackrel{t
    \rightarrow \infty}{\approx} \left\{ \begin{array}{ccc}
      \frac{(4 \pi)^{\frac{d}{2}} \rho_0^2}{|\Gamma(1-d/2)|
      \alpha_C} t^{\frac{d}{2}-1} & \mbox{for} & 2 < d < 4 \\
      \frac{\rho_0^2}{4 A_2 \alpha_C} t & \mbox{for} & d > 4
    \end{array} \right.
    \EEQ
    \renewcommand{\arraystretch}{1.0}
    where $A_2$ is a known constant which is defined in \cite{Paes04a}.
  \item \underline{$\alpha > \alpha_C$ or $d < 2:$}   
    \BEQ
    \label{gl:asymptotic2}
    F(\vec{0},t) \stackrel{t
    \rightarrow \infty}{\approx} A\rho_0^2  \exp(t/\tau_{ts}).
    \EEQ
    The known prefactor $A$ and the time-scale $\tau_{ts}$ are 
    dimension-dependent
    and positive. The exact expressions for them are not essential for our 
    considerations and can be found in \cite{Paes04a}.     
\end{itemize}

The solution of the equations of motion is quite analogous to the one of 
the bosonic contact process
and the results from section~3 can be largely taken over. We find, again
for initially uncorrelated particles of mean density
$\rho_0$,
\BEQ
\label{gl:resultm2}
F(\vec{r};t,s) =  \rho_0^2+ \alpha \int_0^s \!\D\tau\: 
F(\vec{0},\tau) b \left(\vec{r},\demi (t+s)-\tau \right)  
\EEQ
For $t=s$ this formula agrees with 
\cite[eq. (21)]{Paes04a} as it should. We are interested in
the behaviour of the connected correlation function, see (\ref{gl:gdef}),
in the ageing regime. The analysis of eq.~(\ref{gl:resultm2}) is greatly
simplified by recognizing that, quite in analogy with ageing in simple
ferromagnets, there is some intermediate time-scale $t_p$ such that for times
$\tau\lesssim t_p$, one still is in some quasi-stationary regime while
for $\tau\gtrsim t_p$ one goes over into the ageing regime and that
futhermore, the cross-over between these regimes occurs very rapidly
\cite{Zipp00}. We denote by $F_{age}(\vec{0},\tau)$ the asymptotic ageing
form of $F(\vec{0},\tau)$ and write 
\BEA
& &\hspace{-1.3truecm} \int_0^s \!\D \tau\: F(\vec{0},\tau) 
    b(\vec{0},\frac{1}{2}(t+s) - \tau) 
\nonumber \\ 
&\hspace{-1.4truecm}=& \hspace{-1.0truecm}\int_0^{t_p} \!\D \tau\:
F(\vec{0},\tau)b(\vec{0},\demi (t+s)-\tau) +  
s \int_{t_p/s}^1 \!\D v\:  F_{age}(\vec{0},s v) 
                 b(\vec{0},\demi (t+s) - \tau v).
\EEA
We denote the first term of the last line by $C_1(t,s,t_p)$. Since
we expect that $t_p\sim s^{\zeta}$ with $0<\zeta<1$ \cite{Zipp00}, 
we can replace the lower integration limit by $0$ in the 
second integral. This leaves us with the result
\BEQ \label{eq:G}
G(\vec{0};t,s) = C_1(t,s,t_p) + \int_0^s \!\D \tau\: 
F_{age}(\vec{0},\tau) b\left(\vec{0},\demi (t+s)-\tau\right).
\EEQ
On the other hand, we have the following rough estimate 
\BEA
\label{gl:mistake}
|C_1(t,s,t_p)| &\leq& t_p \, \max_{\tau \in [0,t_p]}
\left|F(\vec{0},\tau)b\left(\vec{0},\demi (t+s)-\tau\right)\right| 
\nonumber  \\
&\stackrel{s \gg 1}{\approx}& t_p \max_{\tau \in [0,t_p]}
|F(\vec{0},\tau)|s^{-\frac{d}{2}} \left(4
\pi\left(\frac{1}{2}(t/s+1)
-\frac{t_p}{s}\right)\right)^{-\frac{d}{2}}
\EEA
In the three cases (i) $\alpha<\alpha_C$ and $d>2$, (ii) $\alpha=\alpha_C$ and
$2<d<4$ and (iii) $\alpha=\alpha_C$ and $d>4$ this leads by 
eqs.~(\ref{gl:asymptotic1},\ref{gl:asymptotic3}), respectively, to the
upper bounds $|C_1| \lesssim s^{\zeta-d/2}$, $s^{(\zeta-1)d/2}$ 
and $s^{2 \zeta-d/2}$ which vanish for $s$ large more 
rapidly than $G(\vec{0};t,s)\sim s^{1-d/2}$, 
$s^0$ and $s^{2-d/2}$, respectively and which are derived below. 
Hence $C_1(t,s,t_p)$ is irrelevant for 
the determination of $b$ and the scaling functions and will be dropped in 
what follows. Similarly, because of (\ref{gl:asymptotic2}), 
$C_1(t,s,t_p)$ is non-leading if $\alpha>\alpha_C$. 

We have also checked that for $d >4$ this same result can be derived 
more explicitly using a Laplace transformation, along the lines 
of \cite{Paes04a}. For the sake of brevity, these relatively straightforward
calculations will not be reproduced here \cite{Baum05a}.  

\subsection{Ageing regime: $\vec{r}=0$ and $s,t-s \gg 1$} 

The most interesting cases are $d > 2$ and $\alpha
\leq \alpha_C$, which we will treat first. The asymptotic
expression for $F(\vec{0},t)$ is of the form 
$F_{age}(\vec{0},t)={\cal A} \rho_0^2 t^\xi$,
where $\xi$ and the prefactor $\cal A$ can be read of from
equations (\ref{gl:asymptotic1})-(\ref{gl:asymptotic3}).

We therefore get for the connected autocorrelator
\BEA
\hspace{-2.0truecm} G(\vec{0};t,s) &\hspace{-0.4truecm}=& 
\frac{\alpha \rho_0^2 {\cal A}}{(4\pi)^{\frac{d}{2}}} 
\int_0^s \!\D \tau\: \tau^\xi \left(
\frac{1}{2}(t+s) - \tau \right)^{-\frac{d}{2}}
\nonumber  \\
&\hspace{-0.4truecm}=& 
\frac{\alpha\rho_0^2 {\cal A}}{(\xi+1)(4 \pi)^{\frac{d}{2}}} s^{\xi+1-\frac{d}{2}} 
\left(\frac{1}{2}(y+1)\right)^{-\frac{d}{2}}
{_2F_1}\left(\frac{d}{2},\xi+1;\xi+2;\frac{2}{y+1}\right)
\EEA
where $y ={t}/{s}$ and ${_2F_1}$ is a hypergeometric function. 
We deduce the general form of the scaling function 
\BEQ \label{gl:scalingfunction_gen}
f_G(y) = \frac{\alpha \rho_0^2 {\cal A}}{(\xi+1) (4 \pi)^{\frac{d}{2}}}
\left( \frac{1}{2}(y+1) \right)^{-\frac{d}{2}}
{_2F_1}\left(\frac{d}{2},\xi+1;\xi+2;\frac{2}{y+1}\right).
\EEQ
and the exponents
\BEQ
b = -\xi-1+\frac{d}{2} \qquad \mbox{and} \qquad  \lambda_G = d
\EEQ
and furthermore $z = 2$, see \cite{Paes04a} and below. For the different 
cases we obtain the following explicit expressions:

\begin{itemize}
\item \underline{$\alpha < \alpha_C$ and $d>2$:}
Here we have $\xi = 0$ and the prefactor is ${\cal A} =
-\frac{1}{\alpha'}$. Therefore, we have a value of
\BEQ \label{gl:b63}
b = \frac{d}{2}-1.
\EEQ
The ${_2F_1}$-function can be rewritten with the help of the
relation (9.121,5) from \cite{Grad80}, so that we obtain an
elementary expression for the scaling function:
\BEQ
\label{gl:aging_reg1}
f_G(y) = \frac{\rho_0^2 \alpha}{\alpha' (2\pi)^{\frac{d}{2}}(d-2) }  
\left((y+1)^{-\frac{d}{2}+1} -(y-1)^{-\frac{d}{2}+1}\right). 
\EEQ

\item \underline{$\alpha=\alpha_C$ and $d>2$:}
Here we have $\xi = \frac{d}{2}-1$ and $\xi = 1$ for $2 < d < 4$
and $d > 4$ respectively. This implies for $b$
\BEQ
\label{gl:result_b1}
b = \left\{ \begin{array}{ccc}
  0             & \mbox{for} & 2 < d < 4 \\
  \frac{d}{2}-2 & \mbox{for} & d > 4
  \end{array} \right..
\EEQ
For $2 < d < 4$ the scaling function is (with the prefactor
${\cal A} = \frac{(4 \pi)^{\frac{d}{2}}}{|\Gamma(1-\frac{d}{2})|
\alpha_C}$ \cite{Paes04a})
\BEQ
f_G(y) = \frac{2^{\frac{d}{2}+1} \rho_0^2 }{d
|\Gamma(1-\frac{d}{2})| } (y+1)^{-\frac{d}{2}} {_2F_1}
\left(\frac{d}{2},\frac{d}{2};\frac{d}{2}+1;\frac{2}{y+1}\right).
\EEQ
For $d > 4$, the scaling function (\ref{gl:scalingfunction_gen}) can again be
written as an elementary function with the help of a Gau{\ss} recursion
relation (eq.~(9.137,4) from \cite{Grad80}):
\BEA
\label{gl:scalingfunction_crit}
f_G(y) &=& \frac{\rho_0^2}{4 A_2 (2\pi)^{\frac{d}{2}}(d-2)(d-4)}
\nonumber \\ 
& & \times \left( (y+1)^{-\frac{d}{2}+2} - (y-1)^{-\frac{d}{2}+2} + 
(d-4)(y-1)^{-\frac{d}{2}+1} \right).
\EEA
\item \underline{$\alpha>\alpha_C$ or $d < 2$:}
Due to the exponential behaviour of $F(\vec{0},\tau)$ we do not
have a scaling behaviour in these cases. The 
integrals which enter the calculation are similar to those encountered 
in (\ref{gl:calc_delta}) and where the time-scale $\tau_{ts}$ and
the factor $A$ are defined in eq.~(\ref{gl:asymptotic2}):
\BEQ
\hspace{-1.5truecm}G(\vec{0};t,s)  
= \frac{\alpha\rho_0^2 A\, e^{{t+s}/{(2\tau)}}}{(4 \pi)^{{d}/{2}}}\,
\tau_{ts}^{\frac{d}{2}} \left[\Gamma
\left(-\frac{d}{2}+1,\frac{t+s}{2\tau_{ts}} \right)- \Gamma
\left(-\frac{d}{2}+1,\frac{t-s}{2\tau_{ts}} \right) \right].
\EEQ
Using the asymptotic behaviour of the the Gamma function 
(\ref{gl:asymptotic_gamma}), we see that the leading term in the
scaling limit is
\BEQ
G(\vec{0};t,s) \simeq \frac{\alpha \rho_0^2A}{(2 \pi)^{\frac{d}{2}}} 
(t-s)^{-d/2} \exp\frac{s}{\tau_{ts}} .
\EEQ
In contrast with the other cases treated before, the connected autocorrelator
{\em increases} exponentially with the waiting time $s$.
\end{itemize}
 
\subsection{$\vec{r}$-dependence for $s,t \gg 1$}

\begin{figure}[htb] 
  \vspace{0.5cm}
  \centerline{\epsfxsize=5.0in\epsfclipon\epsfbox
  {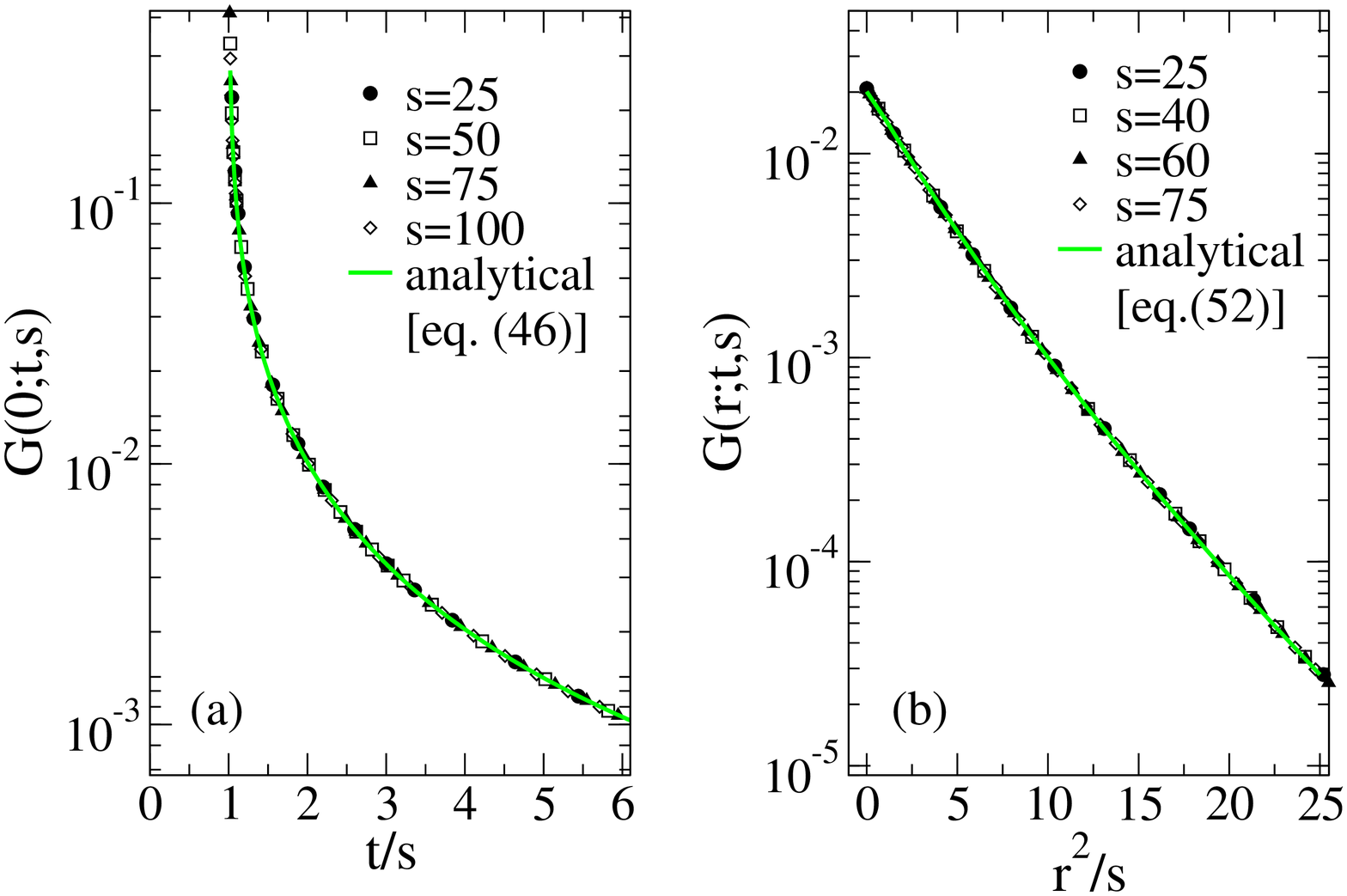}
  }\caption{Scaling behaviour of the two-time correlator $G$ for the case
  $\alpha'=0$ and three dimensions.
  The value $\alpha \rho_0$ was set to unity. In (b), the
  value of $y = {t}/{s}=2$ was used. The data collapse occurs for $b=0$.} 
  \label{fig:p2m2_d3}
\end{figure}

In order to compute the $\vec{r}$-dependence of the correlator, we
follow the same strategy as in the last section. We use
the approximation (\ref{gl:asymptotic_beh2}) which can be justified by an
argument relying on an inequality similar to (\ref{gl:mistake}). 
We obtain the following results.
\begin{itemize}
\item \underline{$\alpha < \alpha_C$ and $d>2$:}
As we have $F(\vec{0},\tau) \approx -\rho_0^2/\alpha'$ the
computation is the same as for the contact process, compare
equation (\ref{gl:resultm1_r}). The result is
\BEQ \label{gl:72}
\hspace{-2.4truecm}
G(\vec{r};t,s) = \frac{-\alpha \rho_0^2}{(4
\pi)^{\frac{d}{2}} \alpha'} \left(\frac{\vec{r}^2}{4}
\right)^{-(\frac{d}{2}-1)} \left[ \Gamma \left(
\frac{d}{2}-1, \frac{\vec{r}^2}{2(t+s)} \right) -\Gamma \left(
\frac{d}{2}-1, \frac{\vec{r}^2}{2(t-s)} \right) \right].
\EEQ
\begin{figure}[ht]
  \vspace{0.5cm}
  \centerline{\epsfxsize=5.0in\epsfclipon\epsfbox
  {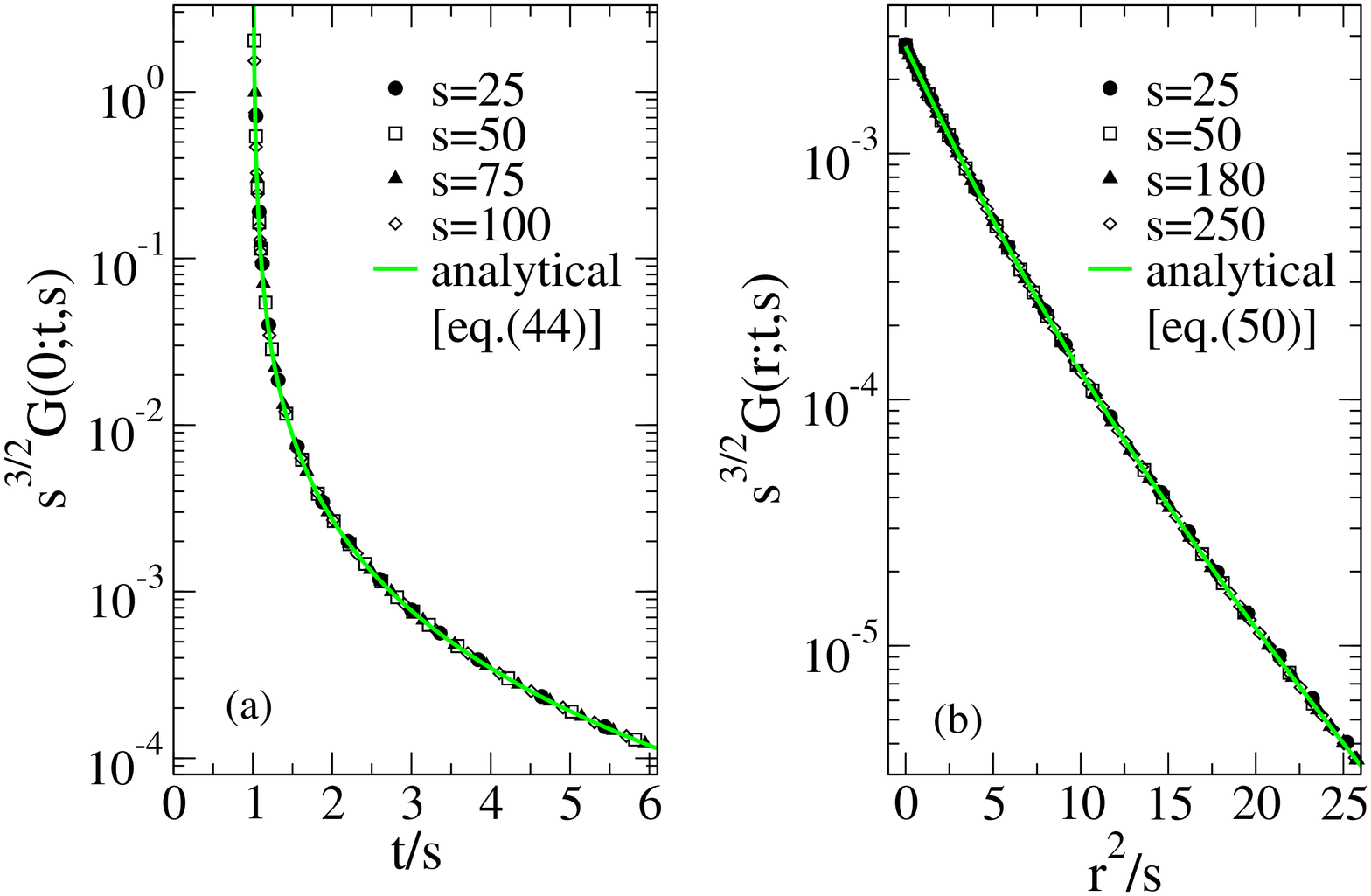}
  }\caption{Scaling plots for the case $\alpha' < 0$ 
  of (a) the autocorrelation function $G(0;t,s)$ and 
  (b) the space-dependent correlation function
  $G(\vec{r};t,s)$ for the bosonic pair-contact process 
  in five dimensions and with $\alpha \rho_0 =
  1$. In (b), the value of $y = {t}/{s}=2$ was used.} 
  \label{fig:p2m2_undercritical}
\end{figure} 
\item \underline{$\alpha = \alpha_C$ and $d > 4$:} We find
  the following result
\BEA
\hspace{-1.0truecm}
G(\vec{r};t,s) &=& 
\frac{\rho_0^2}{4 A_2 (4 \pi)^{\frac{d}{2}}} 
\left( \frac{\vec{r}^2}{4}\right)^{-(\frac{d}{2}-1)}
\nonumber \\ 
&\times &\left[ \frac{t+s}{2} \left(
\Gamma \left(\frac{d}{2}-1,\frac{\vec{r}^2}{2 (t+s)}\right) - 
\Gamma\left(\frac{d}{2}-1,\frac{\vec{r}^2}{2 (t-s)}\right)\right)\right. 
\nonumber \\
& & \hspace{-0.4truecm}-\left.\left(\frac{\vec{r}^2}{4} \right)  \left(
\Gamma \left( \frac{d}{2}-2, \frac{\vec{r}^2}{2 (t+s)}
\right)- \Gamma \left( \frac{d}{2}-2, \frac{\vec{r}^2}{2
(t-s)} \right) \right) \right].
\label{gl:G74}
\EEA
It is straightforward to check consistency with
(\ref{gl:scalingfunction_crit}) for the
case $s$ and $t-s$ much larger than $\vec{r}^2$ by
using the asymptotic form (\ref{gl:asymptotic_gamma}) of the
Gamma function. 
\item \underline{$\alpha = \alpha_C$ and $ 2 < d < 4$:}
\begin{displaymath}
\hspace{-1.9truecm}G(\vec{r};t,s) = \frac{\rho_0^2}{(d/2)
|\Gamma(\frac{d}{2}-1)|} \int_0^s \!\D \tau\, \tau^{\frac{d}{2}-1}
(\demi(t+s) - \tau)^{-\frac{d}{2}} \exp \left(-\frac{\vec{r}^2}{2
(t+s - 2\tau)} \right).
\end{displaymath}
We develop now the exponential function. The integrals 
are similar to those already seen so that we merely state the result
\BEA
\label{gl:75}
\hspace{-1.0truecm}
G(\vec{r};t,s) &=& \frac{2 \rho_0^2}{d
|\Gamma(\frac{d}{2}-1)|} \sum_{n=0}^\infty \frac{1}{n!} \left(
-\frac{\vec{r}^2}{4 s} \right)^n 
\left( \frac{t/s+1}{2} \right)^{-\frac{d}{2}-n} \nonumber \\
& &\times {_2F_1}\left(\frac{d}{2}+n,\frac{d}{2};\frac{d}{2}+1;
\frac{2}{t/s+1} \right).
\EEA
\item \underline{$\alpha > \alpha_C$ or $d  < 2$:}
Here again we develop the exponential function and obtain as final result
\BEA
\label{gl:76}
\hspace{-1.1truecm} G(\vec{r};t,s) &\hspace{-0.1truecm}~=& 
\hspace{-0.0truecm} \frac{\alpha \rho_0^2 A}{(4\pi)^{\frac{d}{2}}}
\sum_{n=0}^\infty \frac{1}{n!} \left(
-\frac{\vec{r}^2}{4} \right)^n
\exp\left({\frac{t+s}{2\tau_{ts}}}\right)
\tau_{ts}^{\frac{d}{2}+n} \nonumber \\
& &\times \left(\Gamma \left(-\frac{d}{2}-n+1,
\frac{t+s}{2\tau_{ts}}\right)-\Gamma \left(-\frac{d}{2}-n+1,
\frac{t-s}{2\tau_{ts}}\right)\right).
\EEA
\end{itemize}
\begin{figure}[htb]
  \vspace{0.5cm}
  \centerline{\epsfxsize=5.0in\epsfclipon\epsfbox
  {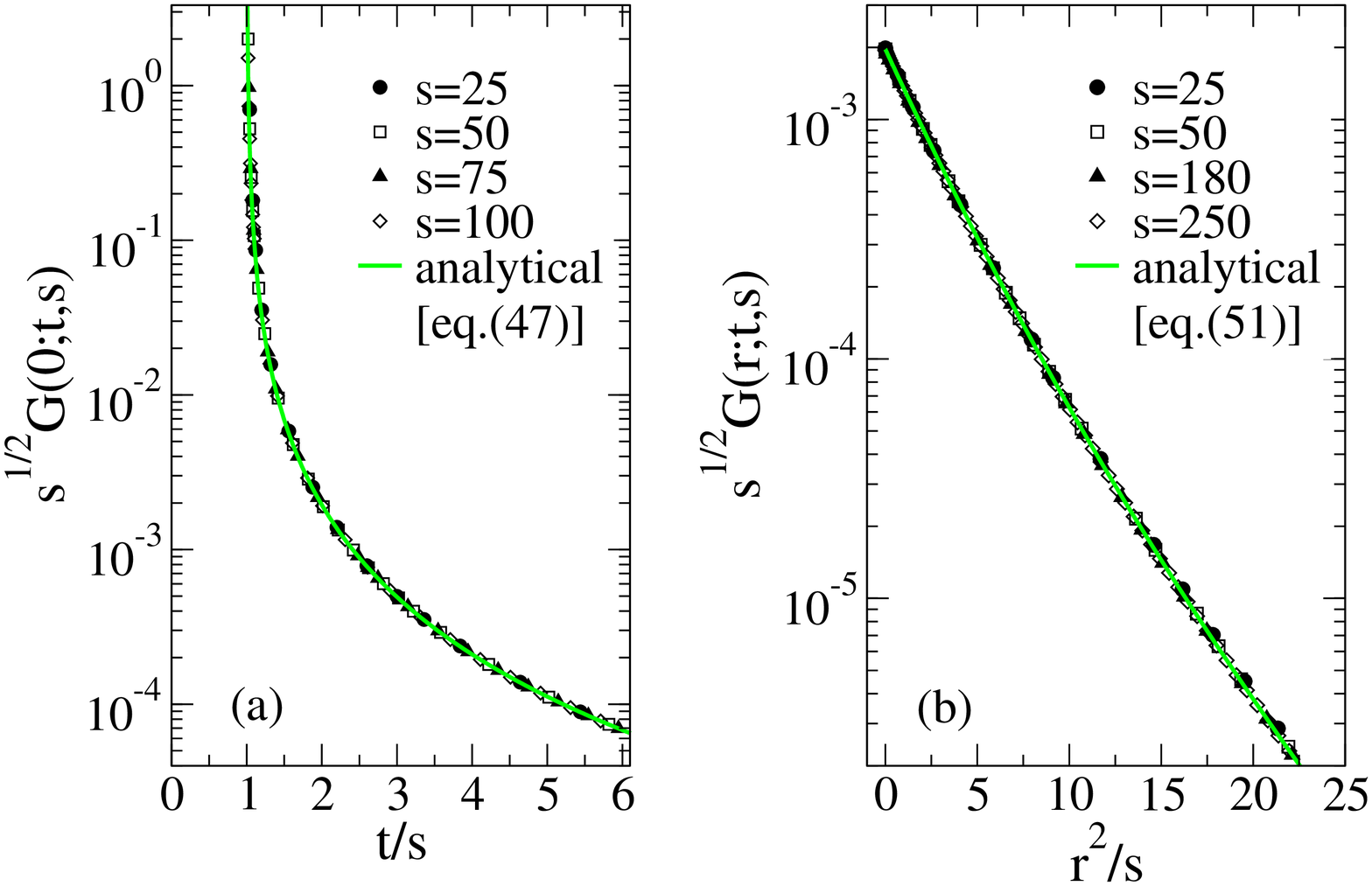}
  }\caption{Scaling plots for the case $\alpha'=0$ of (a)
  the autocorrelation function $G(0;t,s)$ and (b) the
  space-dependent correlation function $G(\vec{r};t,s)$ for
  the bosonic pair-contact process in
  five dimensions and with $\alpha \rho_0 = 1$. In (b), the
  value of $y = {t}/{s}=2$ was used.} 
  \label{fig:p2m2_critical}
\end{figure}
In view of the numerous approximations needed to derive these results,
it is of interest to check them numerically. 
In figure~\ref{fig:p2m2_undercritical}, we compare the results of the
numerical integration of (\ref{gl:resultm2}) with the analytical predictions
(\ref{gl:b63},\ref{gl:aging_reg1},\ref{gl:72}) which apply 
for $\alpha<\alpha_C$ and $d=5$. The nice 
collapse of the data shows that the scaling regime is already reached for the
relatively small values of $s$ used. The perfect agreement of the data with
the analytical results confirms that dropping the term $C_1$ in (\ref{eq:G}) is
justified (and suggests that $C_1$ should be considerably 
smaller than the rough estimate (\ref{gl:mistake})). 
Similarly, we compare data for $\alpha=\alpha_C$ in $5D$ with the
predictions (\ref{gl:scalingfunction_crit},\ref{gl:G74}) in 
figure~\ref{fig:p2m2_critical} and similarly
in $3D$ in figure~\ref{fig:p2m2_d3}. Again the agreement is perfect. 

\section{Response functions} \label{sec:response}
The response function of the first moment to an external
field $h(\vec{y},s)$ is given by
\BEQ
\label{gl:def_response}
R(\vec{x},\vec{y};t,s) := \left.\frac{\delta \langle a(\vec{x},t)
\rangle}{\delta h(\vec{y},s)}\right|_{h=0}.
\EEQ

\subsection{The contact process}
We apply the definition (\ref{gl:def_response}) on both sides of the equation 
of motion (\ref{gl:equation_of_motion_delta}) and find, exploiting 
spatial translation-invariance, with $\vec{r}=\vec{x}-\vec{y}$
\BEQ
\frac{\partial}{\partial t} R(\vec{r};t,s) 
= \frac{1}{2} \Delta R(\vec{r};t,s)
+ \frac{1}{2} \eta R(\vec{r};t,s)  + \delta(t-s).
\EEQ
This is the defining equation of a diffusion-type Green's
function with the solution
\BEQ
\label{gl:resp_exact}
R(\vec{r};t,s) = r_0 \, e^{\frac{1}{2} \eta (t-s)}
b\left(\vec{r},\frac{1}{2}(t-s)\right) \Theta(t-s).
\EEQ
where $b(\vec{r},t)$ was given in eq.~(\ref{gl:bessel}) and $r_0$ is a 
normalization constant. This expression is invariant under time-translations and 
does remain so even at criticality, see below.\footnote{Ageing is characterized 
by the existence of several competing stable stationary states (or a critical 
point) and time-translation invariance ({\sc tti}) can no longer 
be {\em requested}. 
However, that does not mean that {\sc tti} were always impossible and indeed
{\sc tti} can be recovered as a limit case, for certain specific values of
the ageing exponents. A well-known example is the the response function of the 
spherical model quenched onto criticality ($T=T_{\rm c}$) in $d>4$ space 
dimensions \cite{Godr02}.}

\subsection{The critical pair-contact process}
The equation of motion for the particle-density on the
critical line does not change in comparison with the contact process, 
so that we can take over the result (\ref{gl:resp_exact}) with
$\eta$ set to zero and have
\BEQ \label{gl:Rcrit}
R(\vec{r};t,s) = r_0 \,
b\left(\vec{r},\frac{1}{2}(t-s)\right) \Theta(t-s).
\EEQ
The autoresponse function in the scaling regime is obtained
by setting $\vec{r}=0$ and using the known asymptotic behaviour of the
Bessel function, with the result ($t>s$)
\BEQ \label{gl:resp_app}
R(t,s) := R(\vec{0};t,s) \simeq r_0 \,(2 \pi (t-s))^{-{d}/{2}}
\EEQ
from which we can read off the scaling function and the
exponents $a$ and $\lambda_R$
\BEQ
a=\frac{d}{2}-1, \qquad \quad f_R(y) =
\frac{r_0}{(2\pi)^{{d}/{2}}}\, (y-1)^{-{d}/{2}}, \qquad
\quad \lambda_R = d 
\EEQ
\begin{table}[t]
  \[
  \begin{array}{||c|c||c|c||}  \hline \hline
    & & \multicolumn{2}{c||}{\mbox{bosonic pair-contact process}} \\
         \cline{3-4}  
    & \raisebox{1.5ex}[-1.5ex]{ \mbox{bosonic contact process}}
    & \alpha < \alpha_C  & \alpha = \alpha_C  \\ \hline
    \hline
    a  & \frac{d}{2}-1 & \frac{d}{2}-1 & \frac{d}{2}-1  \\
    \hline
    b & \frac{d}{2}-1& \frac{d}{2}-1  & \begin{array}{ccl}
      0 & \mbox{if} & 2 < d < 4  \\
      \frac{d}{2}-2 & \mbox{if} & d > 4 
    \end{array}
      \\
    \hline
     \lambda_R & d & d & d\\ \hline 
     \lambda_G & d & d & d \\ \hline
     z         & 2 & 2 & 2 \\ \hline \hline
  \end{array}
  \]
\caption[AgeingTab1]{Ageing exponents of the critical bosonic contact 
and pair-contact processes in the different regimes. The results for the 
bosonic contact process hold for an arbitrary dimension $d$, but for the
bosonic pair-contact process they only apply if $d>2$, since $\alpha_C=0$ 
for $d\leq 2$.}
\label{tb:results_exp}
\end{table}

We collect our results for the ageing exponents $a,b,\lambda_G,\lambda_R,z$ in
table~\ref{tb:results_exp}. A few comments are now in order. 
First, for both the critical bosonic
contact process and the critical bosonic pair-contact process with
$\alpha<\alpha_C$, we see by comparing the result for $a$ with the 
corresponding ones for $b$, see table~\ref{tb:results_exp}, that $a=b$.
Together with the identity $\lambda_G=\lambda_R$ the critical ageing behaviour
of these systems is quite analogous to the one of simple, reversible
ferromagnets quenched to their critical temperature. Second, the critical
bosonic pair-contact process with $\alpha=\alpha_C$
furnishes an analytically
solved example where $a$ and $b$ are different. This is analogous to the
result found for the $1D$ and $2D$ critical ordinary contact process,
where $a=b-1$ was observed \cite{Enss04,Rama04} and where the relation
$\lambda_G=\lambda_R$ holds as well. However, there is no
apparent simple and general relation between the exponents $a$ and $b$ for
ageing systems without detailed balance. Third, our results for the
critical bosonic pair-contact process provide further evidence against
the generality of a recent proposal by Sastre et {\it al.} \cite{Sast03}
to define a non-equilibrium temperature which was based on the implicit 
assumption that $a=b$ would remain true even in the absence of detailed
balance. Fourth, we can compare the form of the scaling function $f_R(y)$ of the
autoresponse with the prediction of local scale-invariance quoted in
eq.~(\ref{gl:Rf}). We find perfect agreement and identify $a=a'$. Fifth, 
we recall that for $z=2$ there is a variant of local scale-invariance which
takes the presence of a discrete lattice into account. It is possible to
construct the corresponding representation of the Schr\"odinger Lie-algebra and
then a response function transforming covariantly under it should read for
$t>s$ in $d$ spatial dimensions \cite{Henk94b}
\BEQ
\hspace{-1.6truecm} 
R(\vec{r};t,s) = r_0 (t-s)^{(d-2x)/2} \exp\left( \frac{d(t-s)}{\cal M}\right)
{\cal I}_{\vec{r}}\left(\frac{t-s}{\cal M}\right) \;\; , \;\;
{\cal I}_{\vec{r}}(u) := \prod_{j=1}^{d} I_{r_j}(u)
\EEQ
where $x$ is a scaling dimension and $r_0,{\cal M}$ are constants. Here the 
spatial distance $\vec{r}$ is an integer multiple of the lattice constant. 
Comparison with eqs.~(\ref{gl:Rcrit},\ref{gl:bessel}) shows complete agreement
if we identify $x=d/2$ and ${\cal M}=1/2$. 

\section{Conclusions} 
We have studied the ageing behaviour of the exactly solvable
bosonic contact process and of the bosonic critical pair-contact process
in order to get a better understanding on how the present scaling
description of ageing, which is derived from the study of reversible
systems with detailed balance, should be generalized for truly irreversible
systems without detailed balance. This more general situation might be closer
to what is going on in chemical or biological ageing than the reversible
systems undergoing physical ageing, e.g. after a temperature quench.
In comparison with the ordinary contact and pair-contact processes, these
bosonic models permit an accumulation of many particles on a single site and
this possibility does indeed affect the long-time behaviour of these models. 
Trivially, if either particle production or annihilation
dominates, the mean occupation number will either diverge
for large times or the population will die out, but if these
rates are balanced there is a critical line where the
mean particle-density is constant in time and the system's
behaviour is more subtle. Indeed, on the critical line the long-time behaviour 
depends on how effectively single-particle diffusion is capable of homogenizing
the system, see figure~\ref{fig:Abb0}. For dimensions $d\leq 2$, 
there is always {\em clustering} at criticality, 
that is a few sites are highly populated and the others are empty. On the other
hand, for $d>2$ there is no clustering in the bosonic contact process, but in
the bosonic pair-contact process there is a {\em clustering transition}
at some $\alpha=\alpha_C$ such that clustering occurs for $\alpha>\alpha_C$
(where the diffusion is relatively weak) and there is a more or less 
homogeneous state for $\alpha\leq \alpha_C$. 

\begin{table} 
  \[
  \hspace*{-2.5cm}
  \begin{array}{||c|c|c||c|c||}  \hline \hline
    \multicolumn{3}{||c||}{} &  f_R(y) & f_G(y) \\
    \hline
    \hline
    \multicolumn{3}{||c||}{\mbox{contact process}} &
    (y-1)^{-\frac{d}{2}}  & (y-1)^{-\frac{d}{2}+1} -
    (y+1)^{-\frac{d}{2}+1} \\ \hline \hline
    \mbox{pair} & \alpha < \alpha_C & d > 2 & (y-1)^{-\frac{d}{2}} &
    (y-1)^{-\frac{d}{2}+1} - (y+1)^{-\frac{d}{2}+1}  \\ \cline{2-5}
    \mbox{contact} &  & 2 < d < 4 & (y-1)^{-\frac{d}{2}} & 
    (y+1)^{-\frac{d}{2}} {_2F_1}
    \left(\frac{d}{2},\frac{d}{2};\frac{d}{2}+1;\frac{2}{y+1}\right)\\
    \cline{3-5}
    \mbox{process} & \raisebox{1.6ex}[-1.6ex]{$\alpha =
    \alpha_C$} & d > 4 & (y-1)^{-\frac{d}{2}} &  (y+1)^{-\frac{d}{2}+2}
    - (y-1)^{-\frac{d}{2}+2} + ( d-4 ) (y-1)^{-\frac{d}{2}+1} 
\\ \hline \hline
  \end{array}
  \]
\caption{Scaling functions of the autoresponse and autocorrelation 
of the critical bosonic contact and bosonic pair-contact
processes. They are only given up to a multiplicative factor,
which may depend on the dimension. The logarithmic form (\ref{gl:Bos2d}) of
$f_G(y)$ for the $2D$ bosonic contact process may be obtained from a $d\to 2$
limit.}
\label{tb:results_fun}
\end{table}

This behaviour of the models also reflects itself in their ageing behaviour
which we studied here. We anticipated in the ageing regime $t,s\gg 1$ and
$t-s\gg 1$ the scaling forms for the connected
autocorrelator and autoresponse
\BEQ \hspace{-1.5truecm}
G(t,s) := G(\vec{0};t,s) = s^{-b} f_G(t/s) \;\; , \;\;
R(t,s) := R(\vec{0};t,s) = s^{-1-a} f_R(t/s)
\EEQ
together with the asymptotics $f_{G,R}(y)\sim y^{-\lambda_{G,R}/z}$ as $y\gg 1$
and our results for the exponents and the scaling functions are listed in
tables~\ref{tb:results_exp} and \ref{tb:results_fun}.
Specifically:
\begin{enumerate}
\item For $d>2$, the ageing of the bosonic pair-contact process for 
$\alpha<\alpha_C$ lies in the same universality class as the bosonic contact
process, since all critical exponents and the scaling functions co\"{\i}ncide.
Furthermore, the ageing behaviour in the bosonic contact and pair-contact processes
does not depend on whether the parity of the total number of particles is
conserved or not. All these systems have in common that their behaviour is 
strongly influenced by single-particle diffusion. One might wonder whether
an analogy to the Janssen-Grassberger conjecture \cite{Janss81,Grass82}
could be formulated.\footnote{An important ingredient of the models studied
here seems to be that at criticality the mean particle-density stays constant. 
On the other hand, even if a `soft' limit on the particle number per site is
introduced, e.g. by a further reaction $3A\to2A$, the long-time behaviour
is likely to be the one of the PCPD, as checked for the particle-density
in \cite{Park04b}.} 
\item While for $d<2$, we still find a dynamical scaling behaviour in the
critical bosonic contact process, there is no such scaling for the 
bosonic pair-contact process if $\alpha>\alpha_C$, hence in particular 
for $d\leq 2$. Therefore, although both models have the same topology of their 
phase-diagrams for $d<2$, see figure~\ref{fig:Abb0}a, 
their ageing behaviour is different.   
\item At the clustering transition $\alpha=\alpha_C$ in the critical bosonic
pair-contact process, dynamical scaling occurs, but the ageing exponents $a$
and $b$ are different. Here the absence of detailed balance leads to a 
substantial modification of the scaling description with respect to what
happens in critical ferromagnets. In particular, there is no non-trivial 
analogue of the limit fluctuation-dissipation ratio of critical 
ageing ferromagnets. A relation $a\ne b$, see 
(\ref{abverschieden}), has also been observed in the ordinary critical contact
process which also shares the property that $\lambda_G=\lambda_R$ still 
holds \cite{Enss04,Rama04}. However,
according to the known examples, a simple and general relation between 
$a$ and $b$ does not seem to exist
for systems without detailed balance. Further evidence from other 
non-equilbrium models would be welcome. 
\item On the other hand, the equality $\lambda_G=\lambda_R$ between the 
autocorrelation and autoresponse exponents, at the critical point of the
steady-state and for uncorrelated initial states, seems to be a generic feature
even for systems without detailed balance. 
\item The form of the response function is in full agreement with local
scale-invariance which confirms that the annihilation operator $a(\vec{x})$
is a suitable candidate for a quasi-primary field\footnote{In conformal
field-theory, a quasi-primary field transforms covariantly under the 
action of the conformal group \cite{Bela84}. This concept can be generalized to fields
transforming covariantly under the action of a group of local 
scale-transformations, see \cite{Henk02} and references therein for details.} 
of local scale-invariance. 
We shall come back to a detailed analysis of the correlators from the
point of view of local scale-invariance in a sequel paper. 
\end{enumerate}
Explicit results were also derived for the space-dependent scaling functions
of space-time correlator and responses. For the contact process, 
the space-dependent response function is given by eq.~(\ref{gl:resp_exact}) and 
the space-dependent correlation function by eq.~(\ref{gl:resultm1_r}). 
For the critical pair-contact process, the 
space-dependent response function is given by eq.~(\ref{gl:Rcrit}). The space-dependent 
correlation function can be found in (i) eq.~(\ref{gl:72})
for the case $\alpha < \alpha_C$ and $d > 2$, in (ii)
eq.~(\ref{gl:G74}) for the case $\alpha = \alpha_C$ and $d > 4$ , in (iii)  eq.~(\ref{gl:75}) for the case
$\alpha =\alpha_C$ and $2 < d < 4$  and in (iv)
eq.~(\ref{gl:76}) for the case $\alpha > \alpha_C$ or $ d > 2$ . 

Finally, we comment on a suggested relationship between the bosonic
pair-contact process and the spherical model \cite{Paes04a}. In the
spherical model, a classical result by Berlin and Kac \cite{Berl52} states
that the magnetization is spatially uniform, in particular the possibility
that almost the entire macroscopic magnetization were carried by a single
spin can be excluded. This is in remarkable contrast to the clustering 
transition which occurs in the bosonic pair-contact process. More formally,
a closer inspection shows notable differences between the spherical constraint
and the analogous equation used to derive the correlator $F(\vec{0},t)$. 
This suggests that the analogies between the two models
do not seem to have a deeper physical basis. 

{\bf Acknowledgements:} 
This work was supported by the Bayerisch-Franz\"osisches 
Hochschulzentrum (BFHZ).
FB and MP acknowledge the support by the Deutsche Forschungsgemeinschaft 
through grant no. PL 323/2.


\begin{thebibliography}{999}
\bibitem{Stru78} L.C.E. Struik, {\it Physical ageing in amorphous polymers and
other materials}, Elsevier (Amsterdam 1978).
\bibitem{Bray94} A.J. Bray, Adv. Phys. {\bf 43}, 357 (1994). 
\bibitem{Cate00} M.E. Cates and M.R. Evans (eds)
{\it Soft and fragile matter}, IOP Press (Bristol 2000).
\bibitem{Cugl02} L.F. Cugliandolo, in {\it Slow Relaxation and
non equilibrium dynamics in condensed matter}, Les Houches Session 77 July 2002,
J-L Barrat, J Dalibard, J Kurchan, M V Feigel'man eds (Springer, 2003); 
also available at {\tt cond-mat/0210312}.
\bibitem{Godr02} C. Godr\`eche and J.M. Luck, J. Phys.:
  Condens. Matter {\bf 14}, 1589 (2002).
\bibitem{Cris03} A. Crisanti and F. Ritort, J. Phys. A:
  Math. Gen. {\bf 36}, R181 (2003).
\bibitem{Henk04} M. Henkel, Adv. Solid State Phys. {\bf 44}, 389 (2004);
see also {\tt cond-mat/0503739}. 
\bibitem{Cala04} P. Calabrese and A. Gambassi, J. Phys. A:
  Math. Gen. {\bf 38}, R133 (2005).    
\bibitem{Cugl94b} L.F. Cugliandolo, J. Kurchan, and G. Parisi, J. Physique
{\bf I4}, 1641 (1994).
\bibitem{Fish88} D.S. Fisher and D.A. Huse, Phys. Rev. {\bf B38}, 373 (1988).
\bibitem{Huse89} D.A. Huse, Phys. Rev. {\bf B40}, 304 (1989). 
\bibitem{Pico02} A. Picone and M. Henkel, J. Phys. A: Math. Gen. {\bf 35}, 
5575 (2002).
\bibitem{Jans92} H.-K. Janssen, in G. Gy\"orgyi et {\em al.} (eds) {\it
{}From Phase transitions to Chaos}, World Scientific
(Singapour 1992), p. 68.
\bibitem{Henk01} M. Henkel, M. Pleimling, C. Godr\`eche and J.-M. Luck,
Phys. Rev. Lett. {\bf 87}, 265701 (2001).
\bibitem{Henk02} M. Henkel, Nucl. Phys. {\bf B641}, 405 (2002).
\bibitem{Henk05b} M. Henkel and M. Pleimling, J. Phys.: Condens. Matter
{\bf 17}, S1899 (2005).
\bibitem{Henk05a} M. Henkel and M. Pleimling, Europhys. Lett. {\bf 69}, 524 (2005).
\bibitem{Abri04b} S. Abriet and D. Karevski, Eur. Phys. J. {\bf B41}, 79 (2004); {\bf B37}, 47 (2004).
\bibitem{Plei04} M. Pleimling, Phys. Rev. {\bf B70}, 104401 (2004). 
\bibitem{Plei04b} M. Pleimling and A. Gambassi, Phys. Rev.
  {\bf B71}, 180401(R) (2005).
\bibitem{Pico04} A. Picone and M. Henkel, Nucl. Phys. {\bf B688}, 217 (2004).
\bibitem{Henk04b} M. Henkel, A. Picone and M. Pleimling, Europhys. Lett. 
{\bf 68}, 191 (2004).
\bibitem{Enss04} T. Enss, M. Henkel, A. Picone and U. Schollw\"ock,
  J. Phys. A: Math. Gen. {\bf 37}, 10479 (2004).
\bibitem{Rama04} J.J. Ramasco, M. Henkel, M.A. Santos et C.A. da Silva Santos,
J. Phys. A: Math. Gen. {\bf 37} 10497 (2004).
\bibitem{Houc02} B. Houchmandzadeh, Phys. Rev. {\bf E66}, 052902 (2002).
\bibitem{Paes04a} M. Paessens and G.M. Sch\"utz, J. Phys. A: Math. Gen. 
{\bf 37}, 4709 (2004).
\bibitem{Paes04b} M. Paessens, {\tt cond-mat/0406598}.
\bibitem{Henk04c} M. Henkel and H. Hinrichsen, J. Phys. A: Math. Gen. {\bf 37}, R117 (2004).
\bibitem{Howa97} M. Howard and U.C. T\"auber, J. Phys. A:
  Math. Gen. {\bf 30}, 7721 (1997).
\bibitem{Jans04} H.-K. Janssen, F. van Wijland, O. Deloubri\`ere and 
U.C. T\"auber, Phys. Rev. {\bf E70}, 056114 (2004). 
\bibitem{Carl01} E. Carlon, M. Henkel and U. Schollw\"ock, Phys. Rev. {\bf E63},
036101 (2001).
\bibitem{Bark03} G.T. Barkema and E. Carlon, Phys. Rev. {\bf E68}, 
036113 (2003).  
\bibitem{Kock03} J. Kockelkoren and H. Chat\'e, Phys. Rev. Lett. {\bf 90}, 
125701 (2003).
\bibitem{Park04} S.-C. Park and H. Park, Phys. Rev. {\bf E71}, 016137 (2005);
                 Phys. Rev. Lett. {\bf 94}, 065701 (2005).
\bibitem{Szol04} A. Szolnoki, {\tt cond-mat/0408114}.
\bibitem{Hinr05} H. Hinrichsen, {\tt cond-mat/0501075}.
\bibitem{Doi76} M. Doi, J. Phys. A: Math. Gen. {\bf 9}, 1465 and 1479 (1976).
\bibitem{Schu00} G.M. Sch\"utz, in C. Domb and J. Lebowitz (eds) {\it Phase
Transitions and Critical Phenomena}, Vol. 19, London (Acedemic 2000), p. 1.
\bibitem{Abra65} M. Abramowitz and I.A. Stegun, {\it Handbook of mathematical
functions}, Dover (New York 1965). 
\bibitem{Glauber} R.J. Glauber, J. Math. Phys. {\bf 4}, 294
  (1963).
\bibitem{Baum05a} F. Baumann et {\it al.}, {\tt cond-mat/0504243 v1}. 
\bibitem{Grad80} I.S. Gradshteyn and I.M. Ryzhik, {\it Table
  of Integrals, Series, and Products}, 6$^{\rm th}$ edition,
  Academic Press (London 1980). 
\bibitem{Zipp00} W. Zippold, R. K\"uhn and H. Horner, Eur. Phys. J. {\bf B13},
531 (2000).  
\bibitem{Sast03} F. Sastre, I. Dornic and H. Chat\'e, Phys. Rev. Lett.
{\bf 91}, 267205 (2003). 
\bibitem{Henk94b} M. Henkel and G.M. Sch\"utz, Int. J. Mod. Phys. {\bf B8},
3487 (1994).
\bibitem{Janss81} H.K. Janssen, Z. Phys. {\bf B42},151 (1981).
\bibitem{Grass82} P. Grassberger, Z. Phys {\bf B47}, 365 (1982).
\bibitem{Park04b} S.-C. Park, {\tt cond-mat/0412749}.
\bibitem{Bela84} A.A. Belavin, A.M. Polyakov and A.B. Zamolodchikov, Nucl. 
Phys. {\bf B241}, 333 (1984). 
\bibitem{Berl52} T.H. Berlin and M. Kac, Phys. Rev. {\bf 86}, 821 (1952). 
\end{thebibliography}
\end{document}